# Synchronization Behavior of Newton's Cradle

Minseok Lee†[1] Seokchan Hong†[2],

[1]Department of Physics, Korea Advanced Institute of Science and Technology, Korea; danielms.lee@kaist.ac.kr
[2]College of Medicine, Seoul National University, Korea; hong43117@snu.ac.kr
†Contributed equally

A Newton's cradle is a device that demonstrates conservation of momentum using a series of identical colliding pendula. Despite being a famous example that demonstrates the concept of momentum conservation, extensive analysis of the system is rarely reported in literature. Here, we model the system as a collection of identical nonlinear spring pendulums performing viscoelastic collisions, which shows excellent agreement with experiments performed at various conditions. Dependence of its synchronization rate on four key system parameters are studied in detail. Interestingly, the resonance between radial and angular motion was found to modulate the synchronization rate. The proposed theory with full consideration of two dimensional motion and string hysteresis provides an excellent long-term prediction of the synchronized cradle motion.

## 1. INTRODUCTION

Synchronization is a phenomenon where two or more systems interact with each other, resulting in unified motion. The process occurs commonly among system of coupled oscillators. Populations of certain cicada species appear in large numbers every 13 or 17 years [1]. Males of some firefly species in South-East Asia perform synchronized flashing of light [2]. Physical examples include synchronization of metronomes on a common base [3] and voltage oscillations of superconducting Josephson junctions [4-5].

Here, we discuss the synchronization process of the famous Newton's cradle system. When one pendulum at the end is released in a Newton's cradle, it will transfer its momentum by consecutive collisions with adjacent pendulum bobs, finally lifting the one at the other end that repeats the cycle. After some time, they perform synchronized oscillations with slowly decaying amplitude.

Despite its simplicity, the system has been rarely studied extensively. Hutzler *et al.* suggested a model of colliding viscoelastic spheres under harmonic potential and observed a 'beating' phenomenon for imperfect horizontal alignment of pendula [6]. However, the analysis was limited to small-angle and short time regime due to the assumption of harmonic potential, resulting long-term prediction of the motion impossible.

In this paper, the system was modeled as a collection of identical pendulums. A simple linearized model with a single degrees of freedom (angle) specified for each pendulums is first proposed which fails to accurately describe the phenomenon fully. Based on this, a second model is developed with two degrees of freedom (radial and angular) for each pendulum under gravitational potential. As a result, the key processes in synchronization are accurately predicted, with excellent long-term prediction of synchronized motion. Four key system parameters characterizing energy loss through collision, string rigidity, large-angle effects, and horizontal misalignment each are realized, and their effect on the synchronization rate is studied extensively. In particular, resonance between radial and angular motion was found to modulate the synchronization rate. Being a famous example of a piecewise-smooth dynamical system, our work provides a simple interpretation on the Newton's cradle system along with a rigorous experimental verification.

## 2. EXPERIMENTAL METHODS

2.1. Newton's Cradle

The experimental setup was constructed as shown in Fig. 1(a). Table 1 characterizes the carbon steel sphere used for the pendulum bob. Each pendulum bob is connected to the aluminum frame with four Nylon fishing lines that restrict the motion of the bob in a plane by dampening unwanted off-plane motions. The frame acts as a stationary support for the pendulums, minimizing the structural interaction between them. Horizontal alignments were controlled by placing spacers ($\pm 0.01$ mm precision) between two supports. The cradle was initiated by burning a cotton string that holds the end bob in a specified initial condition. To observe the motion of each bob, we shined a high-intensity lamp that is located far away from the cradle that creates point light sources at the focal point of each spheres. Filming with a high-speed camera and applying MATLAB point detection algorithm to the footage allows us to accurately track the motion of the cradle efficiently.

**Table 1. Characterization of the carbon steel pendulum bob**

| Symbol | Quantity | Value |
|---|---|---|
| $R$ | Radius of the bob | 20 mm |
| $E$ | Young's modulus | $200 \times 10^9$ Pa |
| $\nu$ | Poisson's ratio | 0.29 |
| $k_{el}$ | Elastic spring constant | $1.46 \times 10^{10}$ $N/m^{3/2}$ |
| $m$ | Mass of the bob | 0.2813 kg |

2.2. Electric contact resistance experiment

Electric contact resistance experiment was performed to obtain contact information precisely by measuring resistance between each sphere [7]. The circuit was constructed as shown in Fig. 1(c) with specifications of resistor and oscilloscope used in this experiment listed in Table 2. Using binary sequence of resistance connected to the pendulum bobs allows us to figure out which pendulum bobs are in contact. The collision duration can be measured by the length of the signal detected in oscilloscope.

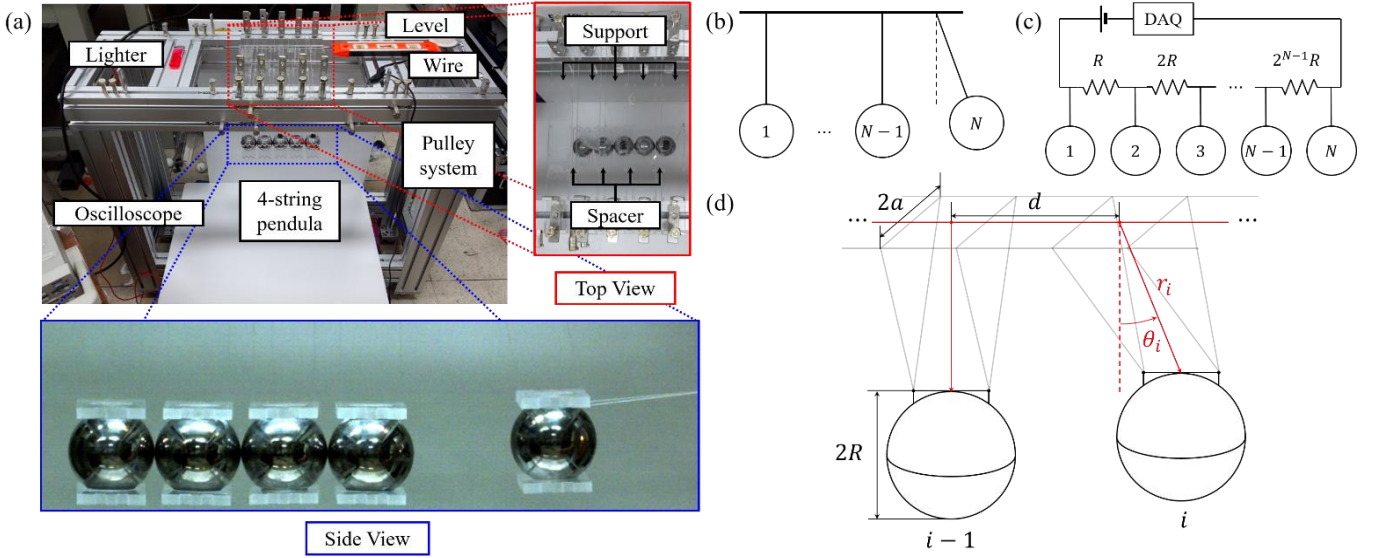

**Fig. 1. Newton's cradle.** (a) Photograph of the experimental set-up. Maximum of five carbon spherical balls were used as pendulum bobs. (b) (c) Circuit diagram of the electric contact resistance experiment. Data acquisition system (DAQ) consists of a resistor ($R_a$) connected parallel to an oscilloscope. (d) Schematic diagram of the four-string cradle. $2a$ is the distance between the pivot points across the pendulum, and $d$ is the distance between adjacent pendula. $r_i$ and $\theta_i$ are the two degrees of freedom specified for the $i$th pendulum, and $2R$ is the diameter of the pendulum bob.

Results show that only the first collision involves more than three bodies, and all the other collisions only involve two bodies. Although there does exist collisions involving more than two bodies near synchronization, the motion of the pendulum bob is not altered significantly due to the small relative velocity between the bobs. Thus, it is safe to assume all the collisions except for the first one as a series of two-body collisions.

**Table 2. Specifications of resistor and oscilloscope**

| Symbol | Quantity | Value |
|---|---|---|
| R | Resistance connected to pendulum bob | 11 Ω |
| $R_a$ | Resistance connected parallel to the oscilloscope | 22 Ω |
| Ω | Sampling rate of oscilloscope | 2.5 G/s |
| M | Bandwidth of oscilloscope | 250 MHz |

## 3. EQUATIONS OF MOTION AND EXPERIMENTAL OBSERVATIONS

3.1 Viscoelastic Collision Model

To simulate collisions between adjacent bobs, we adopt the soft-sphere collision model proposed by Kuwabara and Kono (KK model) [8], where its validity for predicting post collisional velocities and contact duration for viscoelastic spheres are discussed in [9]. The KK model describes the interaction force between two balls $F$ as

$$F = k_{el}\epsilon^{3/2} + k_f \epsilon^{1/2}\dot{\epsilon}, \quad (1)$$

where $\epsilon$ is the 'overlap,' or the decrease in distance between the centers of the two balls, $R$ is the radius of the ball, $k_{el} = \frac{2}{3}\left(\frac{E}{1-\nu^2}\right)\left(\frac{2}{R}\right)^{-1/2}$ is the elastic spring constant, $E$ is the Young's modulus, $\nu$ is the Poisson's ratio, and $k_f$ is the viscoelastic dissipation parameter

The following experiment was employed to measure $k_f$. Newton's cradle system with the three pendulum bobs are prepared. After the first bob is released from some height and collide with the other two bobs, the post-collision velocity of the third bob was measured as a function of the initial velocity of the first bob. The contact duration data was also compared to the KK model in Fig. 2(b). Fitting the KK model to our experimental data with $k_f$ as a fitting parameter, the viscoelastic dissipation parameter was experimentally characterized as $k_f = 18000$ N·s/m$^{3/2}$.

Assuming that the collision takes place instantaneously, we formulate a hard-sphere collision model based on the KK model. The main advantage of the hard-sphere collision model is the utilization of the coefficient of restitution (CoR) concept, which greatly simplifies the collision process. From the KK model, we can obtain CoR as a function of relative velocity at impact. The obtained numerical relation was numerically fitted with the analytic function $e = 1 - \alpha v^n$ where $e$ is CoR, $v$ is relative velocity at impact, $\alpha = 0.0356$, and $n = 0.197$ as shown in Fig. 2(c). This analytic relation describing the CoR as a function of the impact velocity is implemented into our simulations.

3.2. External damping characterization

There are two main sources of external damping which includes aerodynamic drag and pivot friction. Generally, aerodynamic drag differs with the number of pendulum bobs synchronized due to the difference in aerodynamics around the body. Approximating the external damping force $\mathbf{F}_d$ as linear form $\mathbf{F}_d = -b\mathbf{v}$ where $b$ is the external damping coefficient and $\mathbf{v}$ is the velocity vector, we experimentally characterized $b$ as a function of synchronized ball number by fitting the trajectory to a damped sinusoidal function $A \exp\left(-\frac{b}{2m}t\right) \sin(\omega t + \phi)$ – the results are shown in Fig. 2(d). It can be seen that roughly, the external damping coefficient increases with the ball number.

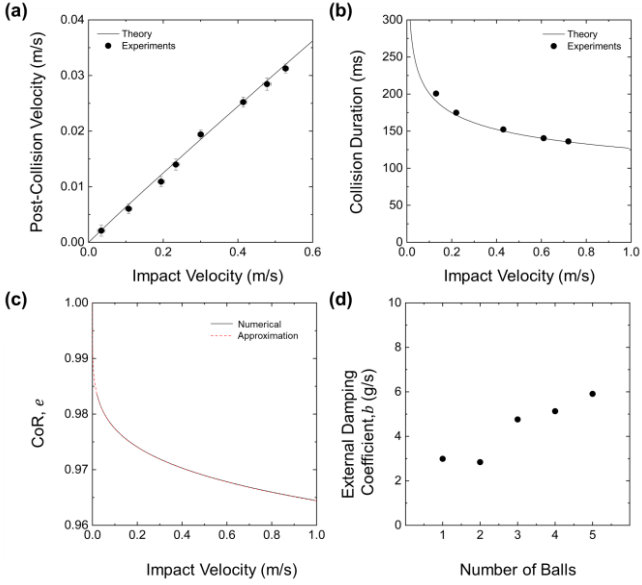

**Fig. 2. Experimental characterization of viscoelastic collision and linear drag coefficient.** (a) Plot of post-collision velocity versus impact velocity of three ball newton's cradle system. The theoretical curve (solid line) is fitted to the experimental data (markers) to characterize $k_f$. (b) Relationship between impact velocity and contact duration for two ball newton's cradle system. The theoretical curve (solid line) is fitted to the experimental data (markers) to characterize $k_f$. (c) Plot of CoR against impact velocity. Numerical result (black solid line) and analytic approximation are shown together. (d) Plot of external damping coefficient against synchronized ball number. Roughly, the external damping coefficient increases with the synchronized ball number.

3.3. Harmonic Oscillator (HO) Model

We first propose a simple model assuming each pendulum as a harmonic oscillator with an angular degree of freedom $\theta_i$ that are coupled with adjacent bobs through inelastic collisions. Assuming that the absolute rotation of each bob is negligible owing to our four-string setup, each pendulum can be properly simplified to a point mass suspended from a pivot. Force balance for $i^{\text{th}}$ pendulum gives

$$m\ddot{\mathbf{r}}_i = -T_i \hat{\mathbf{r}}_i - mg\hat{\mathbf{y}} - b_1 \dot{\mathbf{r}}_i, \quad (2)$$

where $m$ and $\mathbf{r}_i$ are the mass and position vector of the $i^{\text{th}}$ pendulum bob respectively, $g = 9.81$ m/s$^2$ is the gravitational acceleration, $T_i$ is the string tension of the $i^{\text{th}}$ pendulum, and $b_1$ is the external damping coefficient for a single ball. Linearization around the equilibrium point $\theta_i = 0$ of (2) gives the equation of motion in terms of $\theta_i$ as

$$mr_0\ddot{\theta}_i = -mg\theta_i - b_1 r_0 \dot{\theta}_i, \quad (3)$$

where $r_0$ is the pendulum bob's initial radial displacement from the pivot. Trajectories between consecutive collisions are then found by direct time integration of (3) using MATLAB ode45 function. Initial conditions of the system are specified as $\theta_i(0) = 0$ for $i < n$, and $\theta_n(0) = \theta_0$ where $\theta_0 > 0$ is the initial displacement angle of the system. Adjacent bobs are assumed to collide when the distance between their centers are less or equal to their diameter, i.e. $|d + r_0\theta_{i+1} - r_0\theta_i| \leq 2R$.

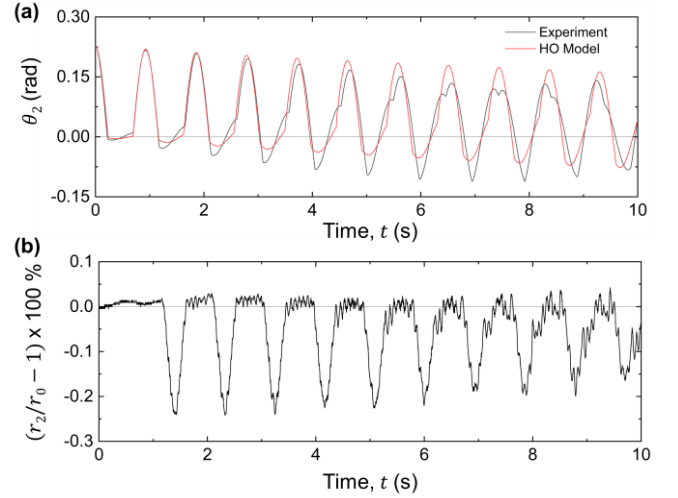

**Fig. 3. Shortcomings of the harmonic oscillator (HO) model.** (a) Comparison of $\theta_2$-time plot under standard parameters for $N = 2$. Theoretical predictions of the HO model (red) shows considerable deviation between our experimental data (black). (b) Stretched percentage of the string connected to the second bob. Large-amplitude oscillation in the radial direction is observed after the second collision.

For cradles with more than three bobs, only the first collision is treated directly using KK model due to the involvement of multiple bodies. All other collisions only involve only two bobs, and hence will be treated using momentum equations with the coefficient of restitution as a function of impact velocity. The resulting velocities after collision are then re-substituted into the equations of motion to be integrated until the next collision occur.

Experimental and simulated trajectories of under standard conditions are shown in the Fig. The simulated trajectory based on the harmonic oscillator model shows considerable deviations from experimental data, particularly on the collision condition. This is due to the assumption of single degree-of-freedom harmonic oscillation. Generally, the pendulum bobs' velocities will not be always parallel to each other during collision. Hence, the collision process must be treated as a two-dimensional collision to accurately predict the collision condition and the post-collision velocity. Also, as period of the pendulum depends on its amplitude in general, the error caused by the constant period assumption will accumulate over time, making the long-term behavior prediction impossible.

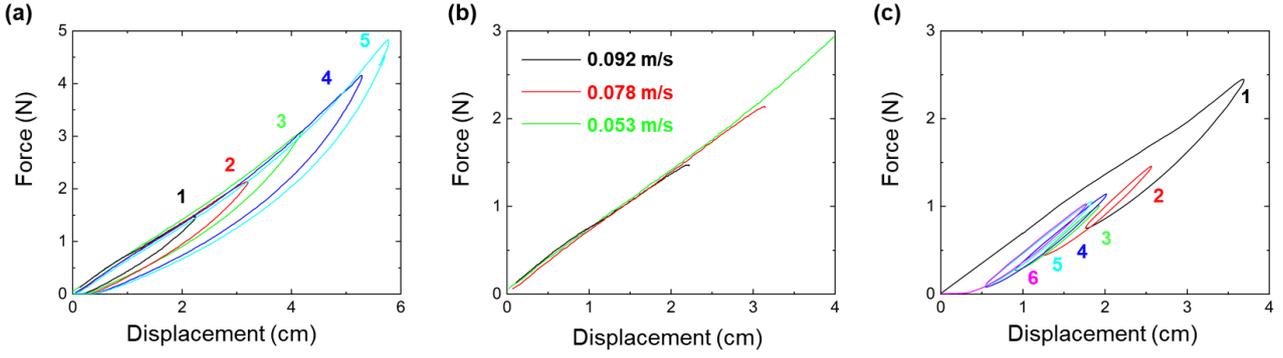

**Fig. 4. Force-displacement curve of the string under various conditions.** (a) Force-displacement curves for a single cycle while varying the maximum displacement/force. Black, red, green, blue and cyan lines correspond to different maximum displacement/force. For regimes associated with the system ($< 3$ N), the loading region can be approximated as linear that is followed by the quadratic unloading-region follows quadratic equation. (b) Force-displacement curve of the string with varying velocity. The Hooke's constant for loading was found to be independent of the loading velocity. (c) Force-displacement curve for multiple cycles. Black, red, green, blue, cyan, and magenta lines correspond to each cycles.

3.4 Nonlinear Spring Pendulum (NSP) Model

Based on the harmonic oscillator model's failure, we propose a second model of a nonlinear spring pendulum. To consider two-dimensional collisions, we must include an additional degree of freedom in the radial direction for each pendulum. Indeed, we were able to observe a significant radial motion during the cradle's operation, justifying the need for the additional degree of freedom – see Fig. 3(b).

The radial degree of freedom facilitates the need to characterize the elastic properties of the string connecting the pendulum bob to the pivot. To characterize the elasticity of the Nylon string used, force-displacement curve of the string was measured under various condition as shown in Fig. 4. For the loading region, it obeys Hooke's law for regimes associated with the system ($< 3$ N). While unloading, the force-displacement curve draws a hysteresis loop. The hysteresis loop is modeled as a linear loading region that is followed by a quadratic unloading region. The string tension operator $T[r, \dot{r}]$ characterizes the hysteresis behavior as

$$T[r,\dot{r}] = \begin{cases} 0 & r < 0 \\ \kappa(r - r_{min}) + T_{min} & r > 0 \text{ and } \dot{r} > 0 \\ T_{max}(r/r_{max})^2 & r > 0 \text{ and } \dot{r} < 0 \end{cases} \quad (4)$$

where $r$ is the displacement of the string from the unstretched length, $\kappa$ is Hooke's constant for loading, and $T_{min/max}$, $r_{min/max}$ are the minimum/maximum tension and displacement of string achieved during the previous unloading/loading region. The equations of motion in terms of the two degrees of freedom $r_i$ and $\theta_i$ finally reads

$$m(\ddot{r}_i - r_i\dot{\theta}_i^2) = -4T_i\left[\sqrt{r_i^2 + a^2} - \sqrt{r_0^2 + a^2}, \dot{r}_i\right]\frac{r_0}{\sqrt{r_0^2+a^2}} + mg\cos\theta_i - b_1\dot{r}_i, \quad (5)$$

$$m(r_i\ddot{\theta}_i + 2\dot{r}_i\dot{\theta}_i) = -mg\sin\theta_i - br_i\dot{\theta}_i, \quad (6)$$

where $2a$ is the distance between the pivot points across the pendulum. Initial conditions of the system are specified as equilibrium positions of each pendulum except for the $n^{th}$ pendulum ($r_n(0) = r_0, \theta_n(0) = \theta_0$). Also, the collision condition is modified to $(d + r_{i+1}\sin\theta_{i+1} - r_i\sin\theta_i)^2 + (r_{i+1}\cos\theta_{i+1} - r_i\cos\theta_i)^2 \leq (2R)^2$. We performed numerical simulations of (5) and (6) as the same manner outlined in 3.3.

Comparing the $\theta_2$ - time plot under standard parameters for $N = 2$, the nonlinear spring pendulum model shows excellent agreement to our experimental data – see Fig.5(a). Also, the cradle is synchronized as the upper and lower amplitude envelope of the pendulums evolve into constant values with opposite sign. We also observe a close match between our theory and experimental data for $N \geq 3$: case of $N = 3$ is shown in Fig.5(b), (c). Other varying parameters yielded great match between theory and experiments.

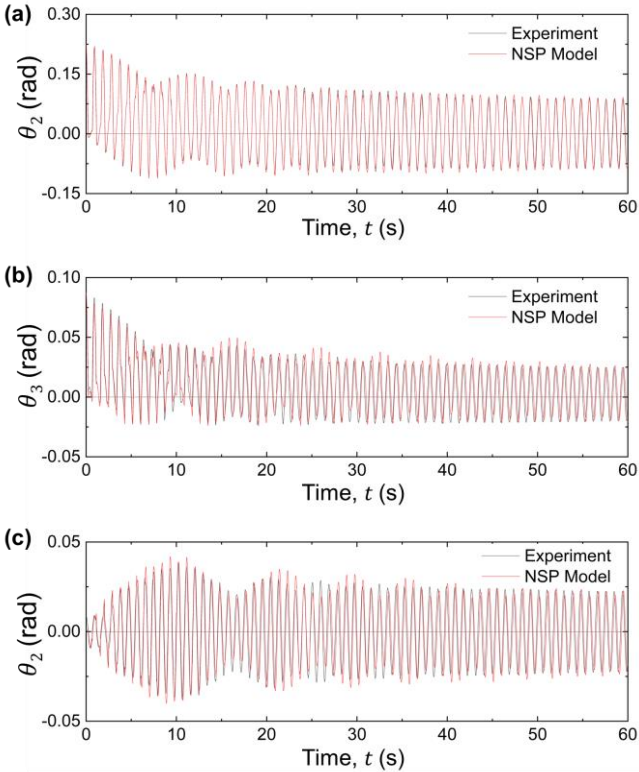

**Fig. 5. Trajectory comparison between experimental data and NSP model under standard parameters.** (a) Comparison of $\theta_2$-time plot for $N = 2$. (b) Comparison of $\theta_3$-time plot for $N = 3$. (c) Comparison of $\theta_2$-time for $N = 3$.

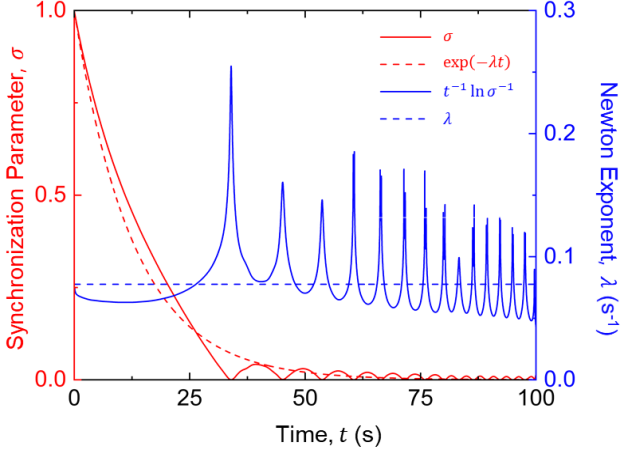

**Fig. 6. Plot of synchronization parameter and Newton exponent against time.** Synchronization parameter (red solid line) is approximated by $\exp(-\lambda t)$ (red dashed line) where $\lambda$ (blue dotted line) is the expected value of $t^{-1}\ln\sigma^{-1}$ (blue solid line).

## 4. SYNCHRONIZATION ANALYSIS

4.1 Synchronization process
The synchronization of Newton's cradle system is achieved when all pendulum bobs move in unison, contacting to each other such that there is no relative motion between each pendulum bob. In an ideal system with no external forces, the total energy after synchronization is exactly $1/N$ of the initial total energy. The total energy of the system can be divided in to potential energy of center of mass (CM), kinetic energy of CM, and sum of the kinetic energy of each ball relative to CM. By momentum conservation, kinetic energy of CM is preserved during collision. Instead, inelastic collision decreases the relative kinetic energy of CM. When we define the initial impact velocity of the $n^{\text{th}}$ ball as $v_0$, the total energy can be divided into two parts: $\frac{1}{2}Nm\left(\frac{V_1}{N}\right)^2$ is the kinetic energy of CM, and $\frac{1}{2}\left(1 - \frac{1}{N}\right)mV_1^2$ is the sum of the kinetic energy of each ball relative to CM. Thus, the total energy at synchronization equals $1/N$ of the initial total energy.

However, in a real system, there exists external forces that dissipate energy too: the string hysteresis and external damping. When two-dimensional collision occurs so that the radial oscillation is induced, energy is dissipated by hysteresis during the string's cycle of elongation and contraction. The energy dissipation by external damping characterized by $b$ was found to be negligible compared to the amount dissipated by collision and string hysteresis, but can be important near the synchronization. Near synchronization, as the pendulum bobs become closer to each other, the external damping coefficient increases as shown in Fig.4(d), accelerating synchronization.

4.2 Synchronization parameter and Newton exponent
The dynamics of the system depends on namely four key parameters $\alpha, \beta, \gamma$, and $\delta$ which are defined as $\alpha = (1-e)v^{-n}$, $\beta = \omega_\theta/\omega_r = \sqrt{mg/\kappa r_0}$, $\gamma = \theta_0^2/6$, $\delta = d/2R - 1$, where $\omega_\theta = \sqrt{g/r_0}$, $\omega_r = \sqrt{\kappa/m}$ are the natural angular frequencies from the linearized angular and radial motion each. Qualitatively, they parametrize energy lost through collision, string rigidity, large-angle effects, and horizontal misalignment respectively. The standard parameters chosen here are $\alpha = 0.0356$, $\beta = 0.114$, $\gamma = 0.00868$, and $\lambda = 0.055$.

Many synchronization phenomena involve the oscillators' phase difference, namely the in-phase and anti-phase synchronization. However, the Newton's cradle system was found to have a single steady synchronous state, where the oscillators are always in-phase throughout the process but their amplitude envelops slowly converge to the same value. Hence, we define the synchronization parameter $\sigma$ as $\sigma = |E_+[\theta_n] + E_-[\theta_n]|/\theta_0$ where $E_\pm[\theta]$ is the upper/lower amplitude envelope of the $\theta$-time graph. Also, to quantify the synchronization rate with a single scalar variable, we define the Newton exponent $\lambda$ as $\lambda = \langle t^{-1}\ln\sigma^{-1}\rangle$ where $\langle \sigma \rangle$ denotes the expected value of $\sigma$. By this definition, $\sigma$ decays from 1 to 0 as synchronization is achieved, and its decay trend is approximated by the exponential function $\exp(-\lambda t)$ – Fig.6 illustrates the definitions of $\sigma$ and $\lambda$ graphically. Fig.7 shows contour plots of $\sigma$ against time and $\alpha, \beta, \gamma, \delta$, and Fig.8 shows plots of $\lambda$ across $\alpha, \beta, \gamma, \delta$. As $\alpha$ is increased, the energy lost per each collision increases accordingly, and $\lambda$ increases accordingly in a monotonic manner. Interestingly, when $\beta$ is increased, an oscillating behavior is observed. This is due to the resonance between the radial and angular oscillations, so that the synchronization rate becomes maximized or minimized depending on $\beta$, or the ratio of angular frequencies in the two motions. This modulation of synchronization rate by resonance may be studied in detail by analyzing the values of $\beta$ corresponding to the peaks in Fig.8(b) in future works. As $\gamma$ is increased, due to the increase in the total energy in the system, the overall impact velocity scales accordingly. However, Fig.2(c) shows that the variation of CoR due to change in impact velocity approaches zero as impact velocity is increased. Therefore, a 'saturation' behavior of $\sigma$ and $\lambda$ on $\gamma$ is expected and illustrated in Fig.7(c) and Fig.8(c). Variation of $\delta$ from the point $\lambda = 0$ regardless of direction produces the 'beating' behavior, where the amplitude envelope of $\theta_n(t)$ converges in an oscillating manner as seen in Fig.7(d), and causes the system to synchronize quicker. However, Fig.8(d) shows that the minimum value of $\lambda$ is attained at around $\delta \sim -1.2 \times 10^{-4}$, implying that the system requires a moderate degree of nonideality for the slowest synchronization.

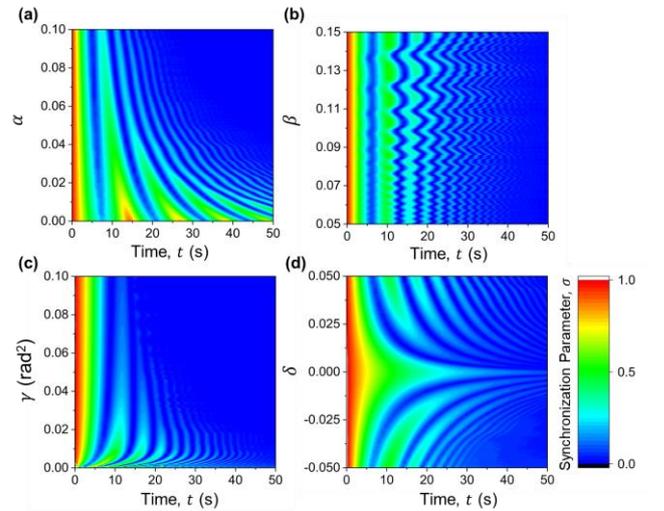

**Fig. 7. Dependence of synchronization parameter on four key system parameters.** (a) Contour plot of $\sigma$ against time and $\alpha$. (b) Contour plot of $\sigma$ against time and $\beta$. Resonance between radial and angular modes are seen by periodic variations along the $\beta$ axis. (c) Contour plot of $\sigma$ against time and $\gamma$. (d) Contour plot of $\sigma$ against time and $\delta$.

## 5. CONCLUSION

This paper considers a dynamic analysis of a Newton's cradle system using a model of $N$ pendulums interacting by viscoelastic collision between adjacent pendulum bobs. A simple model of harmonic oscillators performing viscoelastic collisions with adjacent pendulum bobs (HO model) is derived based on the linearized equations of motion. Observing significant deviation between experimental data and HO model, a modified model of nonlinear spring pendulums (NSP model) is proposed with full consideration into two dimensional motion and two-dimensional collision, which shows excellent long-term experimental agreement across various parameters. A synchronization parameter defined in terms of amplitude envelope is defined to analyze the system's synchronization process in time-domain. Characterization of the decay rate of the synchronization parameter using Newton exponent was performed by varying four key system parameters that characterize energy loss through collision, string rigidity, large-angle effects, and horizontal misalignment each. In particular, the ratio of the natural frequencies in the radial and angular modes of oscillation was found to modulate the synchronization rate by resonance. Also, the system was found to have the minimal synchronization rate when the alignment of pendulums is moderately nonideal. The presented results here provide a simple realization of a piecewise-smooth dynamical system and are expected to be relevant to variety of engineering problems involving collision and hysteresis.

## REFERENCES

[1] Hoppensteadt, F. & Keller, J. Synchronization of periodical cicada emergences. *Science* **194.4262**, 335-337 (1976)
[2] Buck, J. Synchronous rhythmic flashing of fireflies. II. *The Quarterly review of biology* **63.3,** 265-89 (1988)
[3] Pantaleone, J. Synchronization of metronomes. *American Journal of Physics* **70.10**, 992-1000 (2002)
[4] Wiesenfeld, K., Colet, P. & Strogatz, S. H. Synchronization Transitions in a Disordered Josephson Series Array. *Physical Review Letters* **76.3**, 404-407 (1996)
[5] Heath, T. & Wiesenfeld, K. Mutual entrainment of two nonlinear oscillators. *American Journal of Physics* **66.10,** 860-866 (1998)
[6] Hutzler, S., Delaney, G., Weaire, D. & MacLeod, F. Rocking Newton's cradle. *American Journal of Physics* **72.12,** 1508-1516 (2004)
[7] Payr, M. & Glocker, C. EXPERIMENTAL TREATMENT OF MULTIPLE-CONTACT-COLLISIONS. *Energy* **1**, 1 (2005)
[8] Kuwabara, G. & Kono, K. Restitution Coefficient in a Collision between Two Spheres. *Japanese Journal of Applied Physics* **26.8R,** 1230 (1987)
[9] Donahue, C. M., Hrenya, C. M., Zelinskaya, A. P. & Nakagawa, K. J. Newton's cradle undone: Experiments and collision models for the normal collision of three solid spheres. *Physics of Fluids* **20.11**, 3301 (2008)

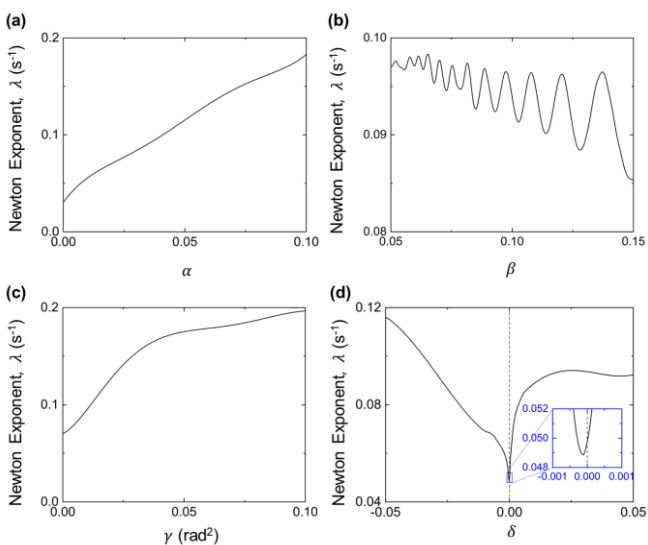

**Fig. 8. Dependence of Newton exponent on four key system parameters.** (a) Plot of $\lambda$ against $\alpha$. (b) Plot of $\lambda$ against $\beta$. Modulation of synchronization rate is visualized by oscillations. (c) Plot of $\lambda$ against $\gamma$. (d) Plot of $\lambda$ against $\delta$. The graph inside represents the zoomed view inside the blue rectangle.